\documentclass[10pt,twocolumn,aps,pra,amsmath,amssymb,showpacs]{revtex4-1}
\usepackage{bm}
\usepackage{mathrsfs}
\usepackage{graphicx}
\usepackage[usenames,dvipsnames,svgnames,table]{xcolor}
\usepackage[unicode=true,
            pdfusetitle, 
            bookmarks=true,
            bookmarksnumbered=false,
            bookmarksopen=false,
            breaklinks=true,
            pdfborder={0 0 0},
            backref=false,
            colorlinks=true]{hyperref}
\hypersetup{linkcolor=NavyBlue,urlcolor=NavyBlue,citecolor=NavyBlue}

\usepackage{amsthm}

\providecommand{\propositionname}{Proposition}

\usepackage{tikz-cd}

\DeclareMathOperator{\tr}{Tr} 			

\newcommand{\cH}{\mathcal{H}}

\begin{document}

\title{Structure of states for which each localized dynamics reduces to a localized subdynamics}
\author{Iman Sargolzahi}
\email{sargolzahi@neyshabur.ac.ir, sargolzahi@gmail.com}
\affiliation{Department of Physics, University of Neyshabur, Neyshabur, Iran}
\author{Sayyed Yahya Mirafzali}
\email{y.mirafzali@vru.ac.ir}
\affiliation{Department of Physics, Faculty of Science, Vali-e-Asr University of Rafsanjan, Rafsanjan, Iran}

\begin{abstract}
We consider a bipartite quantum system $S$ (including parties $A$ and $B$), interacting with an environment $E$ through a \textit{localized} quantum dynamics $\mathcal{F}_{SE}$ . We call a quantum dynamics $\mathcal{F}_{SE}$ localized if, e.g., the party $A$ is isolated from the environment and only $B$ interacts with the environment: $\mathcal{F}_{SE}=id_{A}\otimes \mathcal{F}_{BE}$, where $id_{A}$ is the identity map on the part $A$ and $\mathcal{F}_{BE}$ is a completely positive (CP) map on the both $B$ and $E$.  We will show that the reduced dynamics of the system is also localized as $\mathcal{E}_{S}=id_{A}\otimes \bar{\mathcal{E}}_{B}$, where $\bar{\mathcal{E}}_{B}$ is a CP map on $B$, if and only if the initial state of the system-environment is a \textit{Markov state}. We then generalize this result to the two following cases: when both $A$ and $B$ interact with a same environment, and when each party interacts with its local environment.

\end{abstract}


\maketitle
\section{Introduction}
Consider a quantum system $B$ which undergoes the evolution given by a map $\bar{\mathcal{E}}_{B}$. It is usually argued that $\bar{\mathcal{E}}_{B}$ must be a completely positive (CP) map \cite{1}. This is so because we can always consider another quantum system $A$ which is remained unchanged during the evolution of $B$. So the evolution of the combined system $S=AB$, for arbitrary state  $\rho_{S}=\rho_{AB}$, is given by $id_{A}\otimes \bar{\mathcal{E}}_{B}$, where $id_{A}$ is the identity map on the part $A$. $id_{A}\otimes \bar{\mathcal{E}}_{B}$ must be a positive map (i.e. it must map each positive operator to a positive operator), which means that $\bar{\mathcal{E}}_{B}$ must be a CP map. 

When the system $A$ is remained unchanged, its evolution, obviously, can be represented by $id_{A}$. In the above argument, it is assumed that, in addition, when the evolution of $B$ is given by $\bar{\mathcal{E}}_{B}$, then the evolution of the combined system $S=AB$ is as $id_{A}\otimes \bar{\mathcal{E}}_{B}$. But, as first remarked by Pechukas \cite{21}, there is no reason that this will be the case, in general.

In this paper, we consider the case that only the part $B$ of our bipartite system $S=AB$ interacts with an environment $E$. The evolution of the whole $SE$ is given by $id_{A}\otimes \mathcal{F}_{BE}$, where $\mathcal{F}_{BE}$ is a CP map on the both $B$ and $E$. So, the reduced state of $A$ remains unchanged during the evolution and the reduced dynamics of $A$ can be represented by the identity map $id_{A}$. Now, we question whether the reduced dynamics of $S=AB$ can be represented as $id_{A}\otimes \bar{\mathcal{E}}_{B}$, where $\bar{\mathcal{E}}_{B}$ is a CP map.

Using the results of Refs. \cite{18, 15}, in the next section, we will see that each \textit{localized} dynamics as $id_{A}\otimes \mathcal{F}_{BE}$, for the whole $SE$, reduces to a localized subdynamics as $id_{A}\otimes \bar{\mathcal{E}}_{B}$, if and only if the initial state of  $SE$ be a so-called \textit{Markov state}.

 Therefore, if the initial state of  $SE$ is not a Markov state, there is no guarantee that a dynamics as $id_{A}\otimes \mathcal{F}_{BE}$ reduces to a subdynamics as $id_{A}\otimes \bar{\mathcal{E}}_{B}$. In fact, one can find explicit examples for which localized dynamics does not reduce to localized subdynamics. 
 In other words, one can find explicit examples for which, though the reduced dynamics of $A$ is given by $id_{A}$, but the reduced dynamics of $AB$ can not be represented as $id_{A}\otimes \bar{\mathcal{E}}_{B}$.
Such kind of examples will be given in 
the next section

In our discussion in Sec.~\ref{sec:localized subdynamics}, we use  theorem 1 of Ref.  \cite{15}. During the proof of theorem 1 in Ref. \cite{15}, it is assumed that the final Hilbert spaces, after the evolution, can differ from the initial ones. Whether this assumption can
be relaxed, is discussed in Sec. ~\ref{sec:markov}.

 In Secs.~\ref{sec:both part interaction} and ~\ref{sec:individual environment}, we come back to our main subject and generalize the result of Sec.~\ref{sec:localized subdynamics}. In Sec.~\ref{sec:both part interaction}, we consider the case that the both parts of the system, \textit{A} and \textit{B}, can interact with a same environment and, in Sec.~\ref{sec:individual environment}, we consider the case that each part of the system interacts with its local environment. We end our paper in Sec.~\ref{sec:summary}, with a brief review of our results.

\section{Structure of initial $\rho_{SE}$ for which localized dynamics reduces to localized subdynamics}\label{sec:localized subdynamics}


The quantum dynamics of a finite dimensional system can be written as
\begin{equation}
\label{eq:one}
\begin{aligned}
\rho\rightarrow\rho^{\prime} =\mathcal{F}(\rho)\equiv\sum_{j}F_{j}\,\rho\,F_{j}^{\dagger}, \qquad  \sum_{j}F_{j}^{\dagger}F_{j}=I, 
\end{aligned}
\end{equation}
where $\rho$ and $\rho^{\prime}$ are the initial and final states (density operators) of the system, respectively. $\lbrace F_{j}\rbrace$ is a set of linear operators on $\cH$ ($\cH$ is the Hilbert space of the system) and  $I$ is the identity operator on $\cH$   \cite{1}. Such kind of evolution, given by Eq. (\ref{eq:one}), is called completely positive (CP) evolution \cite{1, 3}. In addition, if the summation in Eq. (\ref{eq:one}) includes only one term, with $F_{1}=U$, for a unitary $U$, then the evolution is called a unitary time evolution; otherwise, the evolution is called a (generalized) measurement.

Assume that the whole system-environment undergoes a CP evolution as Eq. (\ref{eq:one}). In addition, consider the case that the system itself is bipartite $\cH_{S}=\cH_{A}\otimes\cH_{B}$ and the evolution $\mathcal{F}$ in 
Eq. (\ref{eq:one}) is as $\mathcal{F}_{SE}=id_{A}\otimes \mathcal{F}_{BE}$, where $id_{A}$ is the identity map on $\mathcal{L}(\cH_{A})$ and $\mathcal{F}_{BE}$ is a CP map on $\mathcal{L}(\cH_{B}\otimes\cH_{E})$  ($\mathcal{L}(\cH)$ is the space of linear operators on the Hilbert space $\cH$). So the linear operators $F_{j}$ in Eq. (\ref{eq:one}) are in the following form:
\begin{equation}
\label{eq:nineteen}
\begin{aligned}
F_{j}=I_{A}\otimes f_{j},\,\;\; \sum_{j}f_{j}^{\dagger}f_{j}=I_{BE},
\end{aligned}
\end{equation}
where $I_{A}$ ($I_{BE}$) is the identity operator on $\cH_{A}$ ($\cH_{B}\otimes\cH_{E}$) and $f_{j}$ are linear operators acting on $\cH_{B}\otimes\cH_{E}$. We call such a map \textit{localized} since it acts only on 
$BE$. In other words, the party $B$ interacts with the environment $E$ through $\mathcal{F}_{BE}$, but the party $A$ is isolated from the environment and its state remains unchanged during the evolution $\mathcal{F}_{SE}$. Now a naturally arisen question is that whether the reduced dynamics of the system $S$ is also localized.

To find the general structure of initial $\rho_{ABE}$ for which any localized dynamics leads to a localized subdynamics, we first need to recall the defination of \textit{Markov states} \cite{18}.
 For an arbitrary tripartite state  $\rho_{ABE}$, one can find a CP \textit{assignment map} $\Lambda$ such that
$\rho_{ABE}=\Lambda(\rho_{AB})$,
where $\rho_{AB}=\mathrm{Tr_{E}(\rho_{ABE}})$.
For example, $\Lambda$ can be constructed as $\Lambda=\bar{\Lambda} \circ \Xi$. The CP map $\Xi$ is defined as $\Xi(\rho_{AB})=(I_{AB}\otimes\vert0_{E}\rangle)\,\rho_{AB}\,(I_{AB}\otimes\langle 0_{E}\vert)$, where $\vert0_{E}\rangle$ is a fixed state in $\cH_{E}$. The completely positive map $\bar{\Lambda}$, which maps $\rho_{AB}\otimes\vert0_{E}\rangle\langle0_{E}\vert$ to the $\rho_{ABE}$, can be found, e.g., using the method introduced in Ref. \cite{19}. In fact there are infinite number of CP assignment maps $\Lambda$ which map  $\rho_{AB}$ to $\rho_{ABE}$.
However, if one can find a CP assignment map $\Lambda$ as
\begin{equation}
\label{eq:twenty two}
\Lambda=id_{A}\otimes\Lambda_{B},
\end{equation}
i.e. if $\rho_{ABE}=id_{A}\otimes\Lambda_{B}(\rho_{AB})$, where  $\Lambda_{B}:\mathcal{L}(\cH_{B})\rightarrow\mathcal{L}(\cH_{B}\otimes\cH_{E})$ is a CP assignment map on $\mathcal{L}(\cH_{B})$, then the tripartite state $\rho_{ABE}$ is called a Markov state \cite{18}.

 For a Markov state $\rho_{ABE}$, it has been shown in Ref. \cite{18} that there exists a decomposition of the Hilbert space $\cH_{B}$ as $\cH_{B}=\bigoplus_{k}\cH_{b^{L}_{k}}\otimes\cH_{b^{R}_{k}}$  such that
\begin{equation}
\label{eq:twenty three}
\rho_{ABE}=\bigoplus_{k}q_{k}\:\rho_{Ab^{L}_{k}}\otimes\rho_{b^{R}_{k}E},
\end{equation}
where $\lbrace q_{k}\rbrace$ is a probability distribution ($q_{k}\geq 0$, $\sum_{k}q_{k}=1$), $\rho_{Ab^{L}_{k}}$ is a state on $\cH_{A}\otimes\cH_{b^{L}_{k}}$ and $\rho_{b^{R}_{k}E}$ is a state on $\cH_{b^{R}_{k}}\otimes\cH_{E}$. From Eq. (\ref{eq:twenty three}), we see that $\rho_{AB}=\bigoplus_{k}q_{k}\:\rho_{Ab^{L}_{k}}\otimes\rho_{b^{R}_{k}}$, where $\rho_{b^{R}_{k}}=\mathrm{Tr_{E}}(\rho_{b^{R}_{k}E})$. So, the assignment map $\Lambda$ in Eq. (\ref{eq:twenty two}) is as $\Lambda=\bigoplus_{k}id_{Ab^{L}_{k}}\otimes\Lambda_{b^{R}_{k}}$, where $id_{Ab^{L}_{k}}$ is the identity map on $\mathcal{L}(\cH_{A}\otimes\cH_{b^{L}_{k}})$ and $\Lambda_{b^{R}_{k}}:\mathcal{L}(\cH_{b^{R}_{k}})\rightarrow\mathcal{L}(\cH_{b^{R}_{k}}\otimes\cH_{E})$ is a CP assignment map on $\mathcal{L}(\cH_{b^{R}_{k}})$ such that $\Lambda_{b^{R}_{k}}(\rho_{b^{R}_{k}})=\rho_{b^{R}_{k}E}$.

 Using Eqs. (\ref{eq:nineteen}) and (\ref{eq:twenty two}), we have
\begin{equation}
\label{eq:twenty four}
\begin{aligned}
\rho_{AB}^{\prime}=\mathrm{Tr_{E}}\circ \mathcal{F}_{SE}(\rho_{ABE}) \qquad\qquad\qquad\qquad\quad\\
=\mathrm{Tr_{E}}\circ [id_{A}\otimes \mathcal{F}_{BE}]\circ [id_{A}\otimes \Lambda_{B}](\rho_{AB}) \\
=id_{A}\circ [\mathrm{Tr_{E}}\circ \mathcal{F}_{BE}\circ \Lambda_{B}](\rho_{AB}) \qquad\quad\;\;\\
=id_{A}\circ \bar{\mathcal{E}}_{B}(\rho_{AB}),\qquad\qquad\qquad\qquad\qquad
\end{aligned}
\end{equation}
where $\bar{\mathcal{E}}_{B}\equiv\mathrm{Tr_{E}}\circ \mathcal{F}_{BE}\circ \Lambda_{B}$ is a CP map on $\mathcal{L}(\cH_{B})$ (since it is a composition of three CP maps). In addition, $\rho_{AB}=\mathrm{Tr_{E}}(\rho_{ABE})$ and $\rho_{AB}^{\prime}$ are the initial and final states of the system, respectively. So, when the initial $\rho_{ABE}$ is a Markov state, then any arbitrary localized dynamics in Eq. (\ref{eq:nineteen}), leads to a localized subdynamics as Eq. (\ref{eq:twenty four}).

Interestingly, the reverse is also true: if for an initial state $\rho_{ABE}$, any localized dynamics in Eq. (\ref{eq:nineteen}) leads to a localized subdynamics, then $\rho_{ABE}$ is a Markov state. To prove this statement, we need a result of Ref. \cite{15}. Assume that for a tripartite initial state $\rho_{ABE}$ and any arbitrary localized $\mathcal{F}_{SE}=id_{A}\otimes \mathcal{F}_{BE}$ :
\begin{equation}
\label{eq:twenty five}
\begin{aligned}
\rho_{AB^{\prime}E^{\prime}}^{\prime}=\sum_{j}(I_{A}\otimes f_{j})\,\rho_{ABE}\,(I_{A}\otimes f_{j}^{\dagger}),\qquad \\
f_{j}:\cH_{B}\otimes\cH_{E}\rightarrow\cH_{B^{\prime}}\otimes\cH_{E^{\prime}},\qquad 
\sum_{j}f_{j}^{\dagger}f_{j}=I_{BE},
\end{aligned}
\end{equation}
we have
\begin{equation}
\label{eq:twenty six}
\begin{aligned}
\rho_{AB^{\prime}}^{\prime}=\sum_{i}(I_{A}\otimes \bar{E}_{i})\,\rho_{AB}\,(I_{A}\otimes \bar{E}_{i}^{\dagger}),\quad \\
\bar{E}_{i}:\cH_{B}\rightarrow\cH_{B^{\prime}},\qquad 
\sum_{i}\bar{E}_{i}^{\dagger}\bar{E}_{i}=I_{B},\quad
\end{aligned}
\end{equation}
where $\rho_{AB}=\mathrm{Tr_{E}}(\rho_{ABE})$ is the initial state of the system and $\rho_{AB^{\prime}}^{\prime}=\mathrm{Tr_{E^{\prime}}}(\rho_{AB^{\prime}E^{\prime}}^{\prime})$ is the final state of the system. 
In Eqs. (\ref{eq:twenty five}) and (\ref{eq:twenty six}), we assume that, in general, the final Hilbert spaces of the part $B$ and the environment may differ from the initial ones.

The \textit{mutual information} of a bipartite state $\rho_{AB}$ is defined as $I(A:B)_{\rho}=S(\rho_A)+S(\rho_B)-S(\rho_{AB})$, where $\rho_{A}=\mathrm{Tr_{B}}(\rho_{AB})$ and $\rho_{B}=\mathrm{Tr_{A}}(\rho_{AB})$ are the reduced states and $S(\rho)$ is the von Neumann entropy of the state $\rho$: $S(\rho)=-\tr(\rho\log\rho)$ \cite{1}. The mutual information $I(A:B^{\prime})_{\rho^{\prime}}$ of the final state $\rho_{AB^{\prime}}^{\prime}$ is defined similarly. Now, in the theorem 11.15 of Ref. \cite{1}, it has been shown that if Eq. (\ref{eq:twenty six}) holds, then
\begin{equation}
\label{eq:twenty seven}
I(A:B)_{\rho}\geq I(A:B^{\prime})_{\rho^{\prime}}.
\end{equation} 
Since we assume that any arbitrary localized $\mathcal{F}_{SE}=id_{A}\otimes\mathcal{F}_{BE}$ in Eq. (\ref{eq:twenty five}) leads to Eq. (\ref{eq:twenty six}), so for any arbitrary localized $\mathcal{F}_{SE}=id_{A}\otimes \mathcal{F}_{BE}$ in Eq. (\ref{eq:twenty five}), Eq. (\ref{eq:twenty seven}) also holds. Theorem 1 of Ref. \cite{15} states that if for any localized $\mathcal{F}_{SE}=id_{A}\otimes \mathcal{F}_{BE}$ in Eq.~(\ref{eq:twenty five}), the inequality (\ref{eq:twenty seven}) holds, then the initial $\rho_{ABE}$ is a Markov state. So assuming that, for arbitrary localized $\mathcal{F}_{SE}$, Eq.~(\ref{eq:twenty five}) leads to Eq.~(\ref{eq:twenty six}), results that the initial $\rho_{ABE}$ is a Markov state.

In summary, the initial $\rho_{ABE}$ is a Markov state, if and only if, Eq. (\ref{eq:twenty five}) leads to Eq. (\ref{eq:twenty six}), for arbitrary localized $\mathcal{F}_{SE}=id_{A}\otimes \mathcal{F}_{BE}$.
In other words, 

\textbf{Theorem 1.} \textit{If $\rho_{ABE}$  is not a Markov state, then there exists, at least, one CP map $\mathcal{F}_{SE}=id_{A}\otimes \mathcal{F}_{BE}$ which
cannot reduce to a localized subdynamics $id_{A}\circ \bar{\mathcal{E}}_{B}$.}

The above theorem is, in fact, the restatement of (a part of) the theorem 1 of Ref. \cite{15}, in the language appropriate for the case studied in this paper.

Note that, in Eqs. (\ref{eq:twenty five}) and (\ref{eq:twenty six}), we assume that the final Hilbert spaces of the part $B$, $\cH_{B^{\prime}}$, and the environment, $\cH_{E^{\prime}}$, may be different from the initial $\cH_{B}$ and $\cH_{E}$, respectively. In fact, during the proof of the theorem 1 of  Ref. \cite{15}, this assumption has been used. Whether
this assumption can be relaxed, is discussed in the next
section.

We end this section with some illustrating examples.

\textbf{Example 1.} Consider the set $\mathcal{S}=\lbrace\rho_{SE}=\rho_{S}\otimes\tilde{\omega}_{E}\rbrace$,
where $\rho_{S}$ are arbitrary states of the system, but $\tilde{\omega}_{E}$ is a fixed state of environment. As it is famous \citep{1}, when the initial state of the system-environment is a member of  $\mathcal{S}$ and the whole system-environment undergoes a CP evolution as Eq. (\ref{eq:one}), then the reduced dynamics of the system is also CP.

Now, using Eq.~(\ref{eq:twenty three}), it can be shown easily that the factorized initial state $\rho_{ABE}=\rho_{AB}\otimes\tilde{\omega}_{E}$ is a Markov state. It is due to the case that $\cH_{B}=\cH_{b^{L}}\otimes\cH_{b^{R}}$, where $\cH_{b^{R}}$ is a trivial one dimensional Hilbert space. So, for the factorized initial state, any localized dynamics in Eq. (\ref{eq:twenty five}) leads to a localized subdynamics  as Eq.~(\ref{eq:twenty six}). 

\textbf{Example 2.} Consider the set 
\begin{equation}
\label{eq:six}
\begin{aligned}
\mathcal{S}=\lbrace\rho_{SE}=\sum_{i}p_{i}\vert\tilde{i}_{S}\rangle\langle\tilde{i}_{S}\vert\otimes\tilde{\omega}_{i}\rbrace ,
\end{aligned}
\end{equation}
where $\lbrace p_{i}\rbrace$ is arbitrary probability distribution, but $\lbrace\vert\tilde{i}_{S}\rangle\rbrace$ is a fixed orthonormal basis for $\cH_S$ and $\tilde{\omega}_{i}$ are fixed density operators on $\cH_E$. It has been shown in Ref. \cite{11} that when the initial state of the system-environment is a member of  $\mathcal{S}$, then, similar to the factorized initial states in the previous example, the reduced dynamics of the system is CP (for any CP evolution of the whole system-environment). 

But, when the initial $\rho_{ABE}$ is a member of the set given in Eq.~(\ref{eq:six}), then comparing Eqs.~(\ref{eq:six}) and (\ref{eq:twenty three}) shows that, in general, $\rho_{ABE}$ is not a Markov state. For example, consider the case that $\cH_{E}=\bigoplus_{i}\cH_{E_{i}}$ and $\tilde{\omega}_{i}$ in Eq.~(\ref{eq:six}) is a state on $\cH_{E_{i}}$. Let's denote the projector onto the $\cH_{E_{i}}$ by $\Pi_{E_{i}}$. So, from Eq.~(\ref{eq:six}), we have
\begin{equation*}
\mathrm{Tr_{E}}(I_{AB}\otimes\Pi_{E_{i}}\,\rho_{ABE}\,I_{AB}\otimes\Pi_{E_{i}})=p_{i}\,\vert\tilde{i}_{AB}\rangle\langle\tilde{i}_{AB}\vert. 
\end{equation*} 
If $\rho_{ABE}$ can be written as Eq.~(\ref{eq:twenty three}) too, then
\begin{equation*}
\begin{aligned}
\mathrm{Tr_{E}}(I_{AB}\otimes\Pi_{E_{i}}\,\rho_{ABE}\,I_{AB}\otimes\Pi_{E_{i}}) \qquad\\
=\bigoplus_{k}q_{k}\,\rho_{Ab_{k}^{L}}\otimes\mathrm{Tr_{E}}(I_{b_{k}^{R}}\otimes\Pi_{E_{i}}\,\rho_{b_{k}^{R}E}\,I_{b_{k}^{R}}\otimes\Pi_{E_{i}}) \\
=\bigoplus_{k}q_{k}\,\rho_{Ab_{k}^{L}}\otimes q^{(ik)}\,\rho_{b_{k}^{R}}^{(i)},\qquad\qquad\;
\end{aligned} 
\end{equation*} 
where $0\leq q^{(ik)}\leq 1$ and $\rho_{b_{k}^{R}}^{(i)}$ is a state on $\cH_{b_{k}^{R}}$ such that $q^{(ik)}\,\rho_{b_{k}^{R}}^{(i)}=\mathrm{Tr_{E}}(I_{b_{k}^{R}}\otimes\Pi_{E_{i}}\,\rho_{b_{k}^{R}E}\,I_{b_{k}^{R}}\otimes\Pi_{E_{i}})$. So
\begin{equation*}
\bigoplus_{k}q_{k}\,\rho_{Ab_{k}^{L}}\otimes q^{(ik)}\,\rho_{b_{k}^{R}}^{(i)}=p_{i}\,\vert\tilde{i}_{AB}\rangle\langle\tilde{i}_{AB}\vert.
\end{equation*}
Note that $\rho_{Ab_{k}^{L}}\otimes \rho_{b_{k}^{R}}^{(i)}$ belong to different subspaces $\cH_{A}\otimes\cH_{B_{k}}$ ($\cH_{B_{k}}\equiv\cH_{b_{k}^{L}}\otimes\cH_{b_{k}^{R}}$). Therefore the above equality holds with only one term in the summation, i.e. only one $q^{(ik)}$ is non-zero: $q_{k}\,\rho_{Ab_{k}^{L}}\otimes q^{(ik)}\,\rho_{b_{k}^{R}}^{(i)}=p_{i}\,\vert\tilde{i}_{AB}\rangle\langle\tilde{i}_{AB}\vert$. In addition, $\rho_{Ab_{k}^{L}}$ and $\rho_{b_{k}^{R}}^{(i)}$ must be pure states: $\vert\psi_{Ab_{k}^{L}}\rangle\langle\psi_{Ab_{k}^{L}}\vert\otimes\vert\varphi^{(i)}_{b_{k}^{R}}\rangle\langle\varphi^{(i)}_{b_{k}^{R}}\vert=\vert\tilde{i}_{AB}\rangle\langle\tilde{i}_{AB}\vert$. When $\vert\tilde{i}_{AB}\rangle$ is not a separable state as $\vert\psi_{Ab_{k}^{L}}\rangle\otimes\vert\varphi^{(i)}_{b_{k}^{R}}\rangle$, the above equality does not hold. So, in general, the initial $\rho_{ABE}$, chosen from the set in Eq.~(\ref{eq:six}), can not be written as Eq.~(\ref{eq:twenty three}).

Therefore, for set given in  Eq.~(\ref{eq:six}), there is no guarantee that a localized dynamics as Eq.~(\ref{eq:twenty five}) reduces to a localized subdynamics as Eq.~(\ref{eq:twenty six}). In Ref. \cite{7}, we give an explicit example for which a localized dynamics as Eq.~(\ref{eq:twenty five}) does not reduce to a a localized subdynamics as Eq.~(\ref{eq:twenty six}) (for an initial state which can be written as  Eq.~(\ref{eq:six})). This gives an example, illustrating Theorem 1.

\textbf{Example 3.} Consider the set  
\begin{equation}
\label{eq:ten}
\begin{aligned}
\mathcal{S}=\lbrace\rho_{SE}=\bigoplus_{i}p_{i}\,\rho_{L_{i}}\otimes\tilde{\omega}_{R_{i}E}\rbrace ,\quad\\
\cH_S=\bigoplus_{i}\cH_{L_{i}}\otimes\cH_{R_{i}},\;\qquad\quad
\end{aligned}
\end{equation}
where $\lbrace p_{i}\rbrace$ is arbitrary probability distribution, $\rho_{L_{i}}$ is arbitrary state on $\cH_{L_{i}}$, but $\tilde{\omega}_{R_{i}E}$ is a fixed state on $\cH_{R_{i}}\otimes\cH_E$. This set of initial $\rho_{SE}$ also leads to CP reduced dynamics, for arbitrary CP evolution for the whole system-environment \cite{15,16}. In fact, the set given in  Eq.~(\ref{eq:ten}) is the most general possible set of initial $\rho_{SE}$ which leads to CP reduced dynamics, if we restrict ourselves to the case of CP assignment map \cite{16} (see also Ref. \cite{25}).

 When are all $\rho_{ABE}\in S$ in Eq.~(\ref{eq:ten})  Markov states and so for them Eq.~(\ref{eq:twenty five}) leads to  Eq.~(\ref{eq:twenty six}), for any arbitrary localized dynamics $\mathcal{F}_{SE}=id_{A}\otimes \mathcal{F}_{BE}$?

From Eq.~(\ref{eq:ten}) we know that $\cH_{AB}=\bigoplus_{i}\cH_{L_{i}}\otimes\cH_{R_{i}}$. In order that $\rho_{ABE}$ be a Markov state, from Eq.~(\ref{eq:twenty three}), we see that, in addition, we must have $\cH_{B}=\bigoplus_{k}\cH_{b_{k}^{L}}\otimes\cH_{b_{k}^{R}}$. So the simplest case occurs when $\cH_{AB}=\bigoplus_{i}\cH_{A}\otimes\cH_{B_{i}}=\bigoplus_{i}(\cH_{A}\otimes\cH_{b_{i}^{L}})\otimes\cH_{b_{i}^{R}}$, i.e. $\cH_{L_{i}}=\cH_{A}\otimes\cH_{b_{i}^{L}}$ and $\cH_{R_{i}}=\cH_{b_{i}^{R}}$. Now, from Eq.~(\ref{eq:ten}), we have
\begin{equation}
\label{eq:twenty eight}
\rho_{ABE}=\bigoplus_{i}p_{i}\,\rho_{Ab_{i}^{L}}\otimes\tilde{\omega}_{b_{i}^{R}E},
\end{equation} 
which is in the form of Eq.~(\ref{eq:twenty three}) with $\rho_{b_{i}^{R}E}=\tilde{\omega}_{b_{i}^{R}E}$ which are fixed for all $\rho_{ABE}\in S$ in Eq.~(\ref{eq:ten}).

A more general case occurs when $\cH_{A}$ also decomposes as $\cH_{A}=\bigoplus_{i}\cH_{A_{i}}=\bigoplus_{i}\cH_{a_{i}^{L}}\otimes\cH_{a_{i}^{R}}$. So
\begin{equation*}
\begin{aligned}
\cH_{AB}=(\bigoplus_{i}\cH_{A_{i}})\otimes(\bigoplus_{j}\cH_{B_{j}})=\bigoplus_{ij}\cH_{A_{i}}\otimes\cH_{B_{j}} \\
=\bigoplus_{ij}(\cH_{a_{i}^{L}}\otimes\cH_{b_{j}^{L}})\otimes(\cH_{a_{i}^{R}}\otimes\cH_{b_{j}^{R}})\quad\qquad \; \; \\
=\bigoplus_{ij}\cH_{L_{ij}}\otimes\cH_{R_{ij}}.\qquad\qquad\qquad\qquad\qquad\;
\end{aligned} 
\end{equation*}
Then, from Eq. (\ref{eq:ten}), we have
\begin{equation}
\label{eq:twenty nine}
\rho_{ABE}=\bigoplus_{ij}p_{ij}\,\rho_{a_{i}^{L}b_{j}^{L}}\otimes\tilde{\omega}_{a_{i}^{R}b_{j}^{R}E},
\end{equation}
where $\lbrace p_{ij}\rbrace$ is a probability distribution, $\rho_{a_{i}^{L}b_{j}^{L}}$ is a state on $\cH_{a_{i}^{L}}\otimes\cH_{b_{j}^{L}}$ and $\tilde{\omega}_{a_{i}^{R}b_{j}^{R}E}$ is a state on $\cH_{a_{i}^{R}}\otimes\cH_{b_{j}^{R}}\otimes\cH_{E}$. $\tilde{\omega}_{a_{i}^{R}b_{j}^{R}E}$ are fixed for all $\rho_{ABE}\in S$ in Eq.~(\ref{eq:ten}). Now if $\tilde{\omega}_{a_{i}^{R}b_{j}^{R}E}$ be as $\tilde{\omega}_{a_{i}^{R}}^{(j)}\otimes\tilde{\omega}_{b_{j}^{R}E}$ where $\tilde{\omega}_{a_{i}^{R}}^{(j)}$ is a state on $\cH_{a_{i}^{R}}$ and $\tilde{\omega}_{b_{j}^{R}E}$ is a state on $\cH_{b_{j}^{R}}\otimes\cH_{E}$, then Eq.~(\ref{eq:twenty nine}) becomes as Eq.~(\ref{eq:twenty three}). Therefore, for this case, all $\rho_{ABE}\in \mathcal{S}$ in Eq.~(\ref{eq:ten}) are Markov states and so any localized dynamics for them as  Eq. (\ref{eq:twenty five}) reduces to a localized subdynamics as Eq.~(\ref{eq:twenty six}).
    
\textbf{Example 4.} (example 2 of Ref.  \cite{9}) In example 2, we encountered a case for which $\rho_{ABE}\in\mathcal{S}$ in Eq.~(\ref{eq:six}) are not  Markov states. So, for them, according to Theorem 1, one can find at least one localized dynamics as Eq. (\ref{eq:twenty five}) which does not reduce to a localized subdynamics as Eq.~(\ref{eq:twenty six}). There, the initial states of the system $\rho_{AB}=\mathrm{Tr_{E}}(\rho_{ABE})$ are in the restricted form $\sum_{i}p_{i}\vert\tilde{i}_{AB}\rangle\langle\tilde{i}_{AB}\vert$. So, $\rho_{B}=\mathrm{Tr_{A}}(\rho_{AB})=\sum_{i}p_{i}\tilde{\omega}_{B}^{(i)}$, where $\tilde{\omega}_{B}^{(i)}=\mathrm{Tr_{A}}(\vert\tilde{i}_{AB}\rangle\langle\tilde{i}_{AB}\vert)$ are fixed states on $\cH_{B}$, are also restricted. Here, we consider a case for which initial $\rho_{B}$ are arbitrary. 

Assume that the set of initial $\rho_{ABE}$ is given by 
\begin{equation}
\label{eq:seven}
\begin{aligned}
\mathcal{S}=\lbrace\rho_{ABE}=\rho_{B}\otimes\rho_{AE} : \mathrm{Tr_{A}}(\rho_{AE})=\tilde{\omega} \rbrace ,
\end{aligned}
\end{equation}
where $\tilde{\omega}$ is a fixed state. Consider the case that the dynamics is localized as $id_{A}\otimes Ad_{U}$, where $U$ is the swap operator $U\vert\psi\rangle\vert\phi\rangle=\vert\phi\rangle\vert\psi\rangle$ and $Ad_{U}(X)\equiv U X U^{\dagger}$, for arbitrary $X\in \mathcal{L}(\cH_{B}\otimes\cH_{E})$ (note that $U$ acts on $\cH_{B}\otimes\cH_{E}$).

In this example, the reduced dynamics of the part $A$ is given by $id_{A}$. In addition, the reduced dynamics of the part $B$ is given by the CP map $\bar{\mathcal{E}}_{B}(\rho_{B})=\tilde{\omega}$. But, the reduced dynamics of $S=AB$ is not given by $id_{A}\otimes\bar{\mathcal{E}}_{B}$, in general. The final state of $AB$, after the evolution, is given by $\rho_{AE}$, which is the initial state of $AE$. But, we have
\begin{equation*}
id_{A}\otimes\bar{\mathcal{E}}_{B}(\rho_{AB})=id_{A}(\rho_{A})\otimes\bar{\mathcal{E}}_{B}(\rho_{B})=\rho_{A}\otimes\tilde{\omega},
\end{equation*}
where $\rho_{A}= \mathrm{Tr_{E}}(\rho_{AE})$ and so $\rho_{AB}= \mathrm{Tr_{E}}(\rho_{B}\otimes\rho_{AE})=\rho_{A}\otimes\rho_{B}$. In general, $\rho_{AE}$ differs from $\rho_{A}\otimes\tilde{\omega}$, so the reduced dynamics of $AB$ is not given by $id_{A}\otimes\bar{\mathcal{E}}_{B}$.

When $\rho_{AE}=\rho_{A}\otimes\tilde{\omega}$, then the initial $\rho_{ABE}$ is a Markov state as  Eq.~(\ref{eq:twenty three}) and so the reduced dynamics of $AB$ is given by $id_{A}\otimes\bar{\mathcal{E}}_{B}$. But, when $\rho_{AE}\neq\rho_{A}\otimes\tilde{\omega}$, then the initial $\rho_{ABE}$ is not a Markov state and the reduced dynamics of $AB$ is not given by $id_{A}\otimes\bar{\mathcal{E}}_{B}$.

\section{The initial $\rho_{ABE}$ is a Markov state iff each localized dynamics directly reduces to a localized subdynamics} \label{sec:markov}
In Sec.~\ref{sec:localized subdynamics}, we have seen that if each localized dynamics as Eq.~(\ref{eq:twenty five}) leads to a localized subdynamics as Eq.~(\ref{eq:twenty six}), then the initial $\rho_{ABE}$ is a Markov state as Eq.~(\ref{eq:twenty three}) and vice versa. In Eqs.~(\ref{eq:twenty five}) and (\ref{eq:twenty six}), the final $\cH_{B^{'}}$ and $\cH_{E^{'}}$ may differ from the initial $\cH_{B}$ and $\cH_{E}$, respectively. There, we
have questioned whether this condition can be relaxed.
In this section, we discuss about this subject.


Consider a CP assignment map $\Lambda$ which maps $\rho_{AB}$ to $\rho_{ABE}$:
\begin{equation}
\label{eq:B3}
\begin{aligned}
\rho_{ABE}=\Lambda\,(\rho_{AB})
=\sum_{l}R_{l}\,\rho_{AB}\,R^{\dagger}_{l},\\
R_{l}:\;\cH_{A}\otimes\cH_{B}\rightarrow\cH_{A}\otimes\cH_{B}\otimes\cH_{E}, \qquad \\
\sum_{l}R^{\dagger}_{l}R_{l}=I_{AB}. \qquad\qquad
\end{aligned}
\end{equation}
So, for a localized dynamics $\mathcal{F}_{SE}=id_{A}\otimes \mathcal{F}_{BE}$ as 
\begin{equation}
\label{eq:B1}
\begin{aligned}
\rho_{ABE}^{\prime}=\sum_{j}(I_{A}\otimes f_{j})\,\rho_{ABE}\,(I_{A}\otimes f^{\dagger}_{j}), \\
f_{j}:\;\cH_{B}\otimes\cH_{E}\rightarrow\cH_{B}\otimes\cH_{E}, \qquad \\
\sum_{j}f^{\dagger}_{j}f_{j}=I_{BE}, \qquad\qquad
\end{aligned}
\end{equation}
we have
\begin{equation*}
\rho_{ABE}^{\prime}=\sum_{jl}N_{jl}\,\rho_{AB}\,N_{jl}^{\dagger}\;,
\end{equation*}
where $N_{jl}\equiv I_{A}\otimes f_{j}\,R_{l}$ and $\sum_{jl}N_{jl}^{\dagger}N_{jl}=I_{AB}$. Therefore
\begin{equation}
\label{eq:B4}
\begin{aligned}
\rho_{AB}^{\prime}=\sum_{jkl}\langle k_{E}\vert N_{jl}\,\rho_{AB}\,N_{jl}^{\dagger}\vert k_{E}\rangle \\
=\sum_{jkl}X_{jkl}\,\rho_{AB}\,X^{\dagger}_{jkl}\; ,\qquad
\end{aligned}
\end{equation} 
where $\lbrace\vert k_{E}\rangle\rbrace$ is an orthonormal basis for $\cH_{E}$, $X_{jkl}\equiv\langle k_{E}\vert N_{jl}$ is a linear operator on $\cH_{A}\otimes\cH_{B}$ and $\sum_{jkl}X^{\dagger}_{jkl}X_{jkl}=I_{AB}$.
Now if 
\begin{equation}
\label{eq:B22}
\begin{aligned}
X_{jkl}=I_{A}\otimes \bar{E}_{jkl}\; ,\\
\bar{E}_{jkl}:\;\cH_{B}\rightarrow\cH_{B}\; , \\
\sum_{jkl}\bar{E}^{\dagger}_{jkl}\bar{E}_{jkl}=I_{B},
\end{aligned}
\end{equation}  
then
\begin{equation}
\label{eq:B2}
\begin{aligned}
\rho_{AB}^{\prime}=\sum_{jkl}(I_{A}\otimes \bar{E}_{jkl})\,\rho_{AB}\,(I_{A}\otimes \bar{E}_{jkl}^{\dagger}), \\
\bar{E}_{jkl}:\cH_{B}\rightarrow\cH_{B}, \qquad\qquad \\
\sum_{jkl}\bar{E}_{jkl}^{\dagger}\bar{E}_{jkl}=I_{B}; \qquad\qquad
\end{aligned}
\end{equation}
i.e. the reduced dynamics $\mathcal{E}_{AB}$ is also localized: $\mathcal{E}_{AB}=id_{A}\otimes\mathcal{\bar{E}}_{B} $.

If Eq.~(\ref{eq:B22}) holds, we say that the localized dynamics $\mathcal{F}_{SE}=id_{A}\otimes \mathcal{F}_{BE}$ \textit{directly} reduces to the localized sub-dynamics $\mathcal{E}_{AB}=id_{A}\otimes\mathcal{\bar{E}}_{B} $. From Eq.~(\ref{eq:twenty four}), we know that, for a Markov state $\rho_{ABE}$, there exists a CP assignment map as Eq.~(\ref{eq:twenty two}) such that each  $\mathcal{F}_{SE}=id_{A}\otimes \mathcal{F}_{BE}$ directly reduces to a  $\mathcal{E}_{AB}=id_{A}\otimes\mathcal{\bar{E}}_{B} $. In the following, we show that  if for a tripartite state $\rho_{ABE}$ there exists a CP assignment map $\Lambda$ such that a localized dynamics $\mathcal{F}_{SE}=id_{A}\otimes \mathcal{F}_{BE}$ directly reduces to a localized subdynamics   $\mathcal{E}_{AB}=id_{A}\otimes\mathcal{\bar{E}}_{B} $, then this $\Lambda$ is as Eq.~(\ref{eq:twenty two}), i.e. $\rho_{ABE}$ is a Markov state.

Assume that for a CP assignment map $\Lambda$ as  Eq.~(\ref{eq:B3}) and a localized dynamics $\mathcal{F}_{SE}=id_{A}\otimes \mathcal{F}_{BE}$ as Eq.~(\ref{eq:B1}),  Eq.~(\ref{eq:B22}) holds: 
\begin{equation}
\label{eq:B5}
\begin{aligned}
\langle k_{E}\vert I_{A}\otimes f_{j}\,R_{l}=I_{A}\otimes \bar{E}_{jkl}.
\end{aligned}
\end{equation} 
Note that each $R_{l}$ in Eq.~(\ref{eq:B3}) can be decomposed as 
\begin{equation}
\label{eq:B66}
\begin{aligned}
R_{l}=\sum_{m}A_{m}\otimes B_{lm},
\end{aligned}
\end{equation}
where $A_{m}\in \mathcal{L}(\cH_{A})$ and $B_{lm}:\;\cH_{B}\rightarrow\cH_{B}\otimes\cH_{E}$. The set $\mathcal{S}_{A}=\lbrace A_{m}\rbrace$ is chosen as each $A\in \mathcal{L}(\cH_{A})$ can be written as a linear decomposition of $A_{m}$. For example, when $\cH_{A}$ is two dimensional, $\mathcal{S}_{A}$ can be chosen as $\mathcal{S}_{A}=\lbrace I,\,\sigma_{x},\,\sigma_{y},\,\sigma_{z}\rbrace$, where $\sigma_{x}$, $\sigma_{y}$ and $\sigma_{z}$ are the Pauli operators. Note that $A_{m}\in\lbrace I,\,\sigma_{x},\,\sigma_{y},\,\sigma_{z}\rbrace$ are both unitary and orthonormal as 
\begin{equation*}
\mathrm{Tr}(A^{\dagger}_{m}\,A_{n})=2\,\delta_{mn}\;.
\end{equation*}
This way of constructing $\mathcal{S}_{A}$ can be generalized to the higher dimensional cases too. For each finite dimensional $\cH_{A}$ (with the dimension $d_{A}$), one can find a set $\mathcal{S}_{A}=\lbrace A_{m}\rbrace$, including ${(d_{A})}^{2}$ unitary operators, as a basis for $\mathcal{L}(\cH_{A})$, where $A_{1}=I_{A}$ and $A_{m}$ are orthonormal as
\begin{equation}
\label{eq:B7}
\mathrm{Tr}(A^{\dagger}_{m}\,A_{n})=d_{A}\,\delta_{mn}\;;
\end{equation} 
see, e.g., the appendix B of Ref.~\cite{22}.

Replacing  (\ref{eq:B66}) in (\ref{eq:B5}), we get
\begin{equation*}
\sum_{m}A_{m}\otimes\langle k_{E}\vert f_{j}\,B_{lm}=I_{A}\otimes \bar{E}_{jkl}.
\end{equation*}
So, using Eq.~(\ref{eq:B7}), we see that $\bar{E}_{jkl}=\langle k_{E}\vert f_{j}\,B_{l1}$ and, for $m\neq 1$, we have
\begin{equation}
\label{eq:B8}
\langle k_{E}\vert f_{j}\,B_{lm}=0\,.
\end{equation}
Choosing $\lbrace\vert n_{B}\rangle\rbrace$ as an orthonormal basis for $\cH_{B}$, the above equation means that for $m\neq 1$:
\begin{equation}
\label{eq:B9}
\begin{aligned}
\langle k_{E}\vert\langle n^{\prime}_{B}\vert f_{j}\,B_{lm}\vert n_{B}\rangle=0 \\
\Rightarrow\; \langle\varphi_{jkn^{'}}\vert\psi_{lmn}\rangle=0\, ,
\end{aligned}
\end{equation} 
where $\vert\psi_{lmn}\rangle\equiv B_{lm}\vert n_{B}\rangle$ and $\vert\varphi_{jkn^{'}}\rangle\equiv f^{\dagger}_{j}\vert n^{\prime}_{B}\rangle\vert k_{E}\rangle$.
Note that, using  Eq.~(\ref{eq:B1}), $\sum_{jkn^{\prime}} \vert\varphi_{jkn^{'}}\rangle\langle\varphi_{jkn^{'}}\vert=I_{BE}$. So,  for $m\neq 1$, from Eq.~(\ref{eq:B9}),  we have
\begin{equation}
\label{eq:B99}
\begin{aligned}
\langle \psi_{lmn}\vert\psi_{lmn}\rangle=\sum_{jkn^{\prime}}\langle \psi_{lmn}\vert\varphi_{jkn^{'}}\rangle \langle\varphi_{jkn^{'}}\vert\psi_{lmn}\rangle=0. 
\end{aligned}
\end{equation} 
 Therefore, for $m\neq 1$ (and all $n$), $\vert\psi_{lmn}\rangle=0$; so, for $m\neq 1$, $B_{lm}=0$. Finally, $R_{l}$ in Eq.~(\ref{eq:B66}) is as 
\begin{equation}
\label{eq:B10}
R_{l}=I_{A}\otimes B_{l1}\,,
\end{equation}
which means that the CP assignment map $\Lambda$ in Eq.~(\ref{eq:B3}) is as Eq.~(\ref{eq:twenty two}). So, $\rho_{ABE}$ is a Markov state and, using the result of Ref.~\cite{18}, can be written as Eq.~(\ref{eq:twenty three}).

In summary, we have shown that, if  for a localized dynamics $\mathcal{F}_{SE}=id_{A}\otimes \mathcal{F}_{BE}$ as Eq.~(\ref{eq:B1}),  Eq.~(\ref{eq:B22}) holds, then $\rho_{ABE}$ is a Markov state and so any other localized dynamics as Eq.~(\ref{eq:B1}) also  \textit{directly} reduces to a localized subdynamics as Eq.~(\ref{eq:B2}). 
In other words,

\textbf{Theorem 2.}\textit{ If $\rho_{ABE}$ is not a Markov state, then} any arbitrary \textit{localized dynamics as Eq.~(\ref{eq:B1}) cannot} directly \textit{reduce to a localized subdynamics  as Eq.~(\ref{eq:B2}).}


Note that,  when $\rho_{ABE}$ is not a Markov state, the above theorem does not  guarantee that the reduction of a localized dynamics as Eq.~(\ref{eq:B1}) is not  equivalent to any localized subdynamics. Theorem 2 only states that the \textit{direct} reduction of any  localized dynamics is not localized (when $\rho_{ABE}$ is not a Markov state).
 From Theorem 1, we know that, for such a state, there exists, at least, one localized dynamics as Eq.~(\ref{eq:twenty five}) which its reduction is not equivalent to any  localized subdynamics as Eq.~(\ref{eq:twenty six}).  Now, when the initial $\rho_{ABE}$ is not a Markov state, whether it is always possible to find a localized dynamics as Eq.~(\ref{eq:B1}) which its reduction is not equivalent to any localized subdynamics, remains as an open question.

\section{When both parts of the system can interact with the environment}\label{sec:both part interaction}
Let's come back to our main subject. In Sec.~\ref{sec:localized subdynamics}, we considered the case that the system is bipartite $\cH_{S}=\cH_{A}\otimes\cH_{B}$. Then, assuming that only the party $B$ interacts with the environment (and the party $A$ is isolated from the environment) as  Eq.~(\ref{eq:twenty five}), we asked when any such localized dynamics reduces to a localized sub-dynamics as Eq.~(\ref{eq:twenty six}). We have seen that this will be the case if and only if the initial $\rho_{ABE}$ is a Markov state as Eq.~(\ref{eq:twenty three}).  

Now, assume that, also, the party $A$ can interact with the environment. So each localized dynamics as $\mathcal{F}_{SE}=id_{B}\otimes\mathcal{F}_{AE}$ leads to a localized subdynamics as $\mathcal{E}_{AB}=id_{B}\otimes \bar{\mathcal{E}}_{A}$ if and only if the initial $\rho_{ABE}$ be a Markov state as 
\begin{equation}
\label{eq:thirty five}
\begin{aligned}
\rho_{ABE}=\bigoplus_{j}p_{j}\,\rho_{a_{j}^{L}B}\otimes\,\rho_{a_{j}^{R}E}\,, \quad\\
\cH_{A}=\bigoplus_{j}\cH_{A_{j}}=\bigoplus_{j}\cH_{a_{j}^{L}}\otimes\cH_{a_{j}^{R}}\,,
\end{aligned}
\end{equation}
where $\lbrace p_{j}\rbrace$ is a probability distribution, $\rho_{a_{j}^{L}B}$ is a state on $\cH_{a_{j}^{L}}\otimes\cH_{B}$ and $\rho_{a_{j}^{R}E}$ is a state on $\cH_{a_{j}^{R}}\otimes\cH_{E}$.

If we require that arbitrary localized maps $id_{A}\otimes \mathcal{F}_{BE}$ and $id_{B}\otimes \mathcal{F}_{AE}$ reduce as  $id_{A}\otimes \bar{\mathcal{E}}_{B}$ and $id_{B}\otimes \bar{\mathcal{E}}_{A}$, respectively, then both Eqs.~(\ref{eq:twenty three}) and (\ref{eq:thirty five}) must be held simultaneously for the initial $\rho_{ABE}$. Now, consider the projection
\begin{equation}
\label{eq:thirty six}
\begin{aligned}
\Pi_{jk}\equiv\Pi_{A_{j}}\otimes\Pi_{B_{k}}\otimes I_{E} \qquad\qquad\qquad\qquad\\
=(\Pi_{a_{j}^{L}}\otimes\Pi_{a_{j}^{R}})\otimes(\Pi_{b_{k}^{L}}\otimes\Pi_{b_{k}^{R}})\otimes I_{E}\,,
\end{aligned}
\end{equation}
where $\Pi_{A_{j}}$, $\Pi_{B_{k}}$, $\Pi_{a_{j}^{L}}$, $\Pi_{a_{j}^{R}}$, $\Pi_{b_{k}^{L}}$ and $\Pi_{b_{k}^{R}}$ are the projectors onto $\cH_{A_{j}}$, $\cH_{B_{k}}$, $\cH_{a_{j}^{L}}$, $\cH_{a_{j}^{R}}$, $\cH_{b_{k}^{L}}$ and $\cH_{b_{k}^{R}}$, respectively, and $I_{E}$ is the identity operator on $\cH_{E}$. From Eqs.~(\ref{eq:twenty three}), (\ref{eq:thirty five})  and (\ref{eq:thirty six}), we have
\begin{equation}
\label{eq:thirty seven}
\begin{aligned}
\Pi_{jk}\rho_{ABE}\Pi_{jk}=p_{j}\,\sigma_{a_{j}^{L}B_{k}}\otimes\rho_{a_{j}^{R}E} \\
=q_{k}\,\sigma_{A_{j}b_{k}^{L}}\otimes\rho_{b_{k}^{R}E}\, ,
\end{aligned}
\end{equation}
where $\sigma_{a_{j}^{L}B_{k}}\equiv\Pi_{B_{k}}\rho_{a_{j}^{L}B}\Pi_{B_{k}}$ and $\sigma_{A_{j}b_{k}^{L}}\equiv\Pi_{A_{j}}\rho_{Ab_{k}^{L}}\Pi_{A_{j}}$ are positive operators.

Note that if $\sigma_{a_{j}^{L}B_{k}}$ be non-zero then $\sigma_{A_{j}b_{k}^{L}}$ is so, and vice versa (we consider only those terms in Eqs.~(\ref{eq:twenty three}) and (\ref{eq:thirty five}) for which $q_{k}\neq 0$ and $p_{j}\neq 0$). For each $j$ there is at least one $k$ for which $\sigma_{a_{j}^{L}B_{k}}$ and $\sigma_{A_{j}b_{k}^{L}}$ are non-zero. For this $(j, k)$, we define
\begin{equation*}
p^{(j, k)}\equiv\mathrm{Tr}(\sigma_{a_{j}^{L}B_{k}})\, ,\qquad\rho_{a_{j}^{L}B_{k}}^{(j, k)}\equiv\dfrac{\sigma_{a_{j}^{L}B_{k}}}{p^{(j, k)}}\, ,
\end{equation*}
and
\begin{equation*}
q^{(j, k)}\equiv\mathrm{Tr}(\sigma_{A_{j}b_{k}^{L}})\, ,\qquad\rho_{A_{j}b_{k}^{L}}^{(j, k)}\equiv\dfrac{\sigma_{A_{j}b_{k}^{L}}}{q^{(j, k)}}\, ,
\end{equation*}
where $\rho_{a_{j}^{L}B_{k}}^{(j, k)}$ is a state on $\cH_{a_{j}^{L}}\otimes\cH_{B_{k}}$ and $\rho_{A_{j}b_{k}^{L}}^{(j, k)}$ is a state on $\cH_{A_{j}}\otimes\cH_{b_{k}^{L}}$. So, for this $(j, k)$, Eq.~(\ref{eq:thirty seven}) can be rewritten as
\begin{equation*}
p_{j}\,p^{(j, k)}\,\rho_{a_{j}^{L}B_{k}}^{(j, k)}\otimes\rho_{a_{j}^{R}E}=
q_{k}\,q^{(j, k)}\,\rho_{A_{j}b_{k}^{L}}^{(j, k)}\otimes\rho_{b_{k}^{R}E}\,.
\end{equation*}
By tracing from both sides, we get $p_{j}\,p^{(j, k)}=q_{k}\,q^{(j, k)}$. So
\begin{equation}
\label{eq:thirty eight}
\rho_{a_{j}^{L}B_{k}}^{(j, k)}\otimes\rho_{a_{j}^{R}E}=
\rho_{A_{j}b_{k}^{L}}^{(j, k)}\otimes\rho_{b_{k}^{R}E}\,.
\end{equation}
Tracing from both sides, with respect to $a_{j}^{L}$ and $B_{k}$, gives us
\begin{equation}
\label{eq:thirty nine}
\rho_{a_{j}^{R}\,E}=\rho_{a_{j}^{R}}^{(j, k)}\otimes\bar{\rho}_{E}^{(k)},
\end{equation}
where $\rho_{a_{j}^{R}}^{(j, k)}=\mathrm{Tr_{\mathit{a_{j}^{L}\,b_{k}^{L}}}}(\rho_{A_{j}b_{k}^{L}}^{(j, k)})$ is a state on $\cH_{a_{j}^{R}}$ and $\bar{\rho}_{E}^{(k)}=\mathrm{Tr_{\mathit{b_{k}^{R}}}}(\rho_{b_{k}^{R}E})$ is a state on $\cH_{E}$. Similarly, by tracing from both sides of Eq.~(\ref{eq:thirty eight}) with respect to $A_{j}$ and $b_{k}^{L}$, we have
\begin{equation}
\label{eq:forty}
\rho_{b_{k}^{R}\,E}=\rho_{b_{k}^{R}}^{(j, k)}\otimes\hat{\bar{\rho}}_{E}^{(j)},
\end{equation}
where $\rho_{b_{k}^{R}}^{(j, k)}=\mathrm{Tr_{\mathit{a_{j}^{L}\,b_{k}^{L}}}}(\rho_{a_{j}^{L}B_{k}}^{(j, k)})$ is a state on $\cH_{b_{k}^{R}}$ and $\hat{\bar{\rho}}_{E}^{(j)}=\mathrm{Tr_{\mathit{a_{j}^{R}}}}(\rho_{a_{j}^{R}E})$ is a state on $\cH_{E}$.

Using Eq.~(\ref{eq:thirty nine}), we can rewrite Eq.~(\ref{eq:thirty five}) as
\begin{equation*}
\rho_{ABE}=\bigoplus_{j}p_{j}\,\rho_{a_{j}^{L}B}\otimes\rho_{a_{j}^{R}}^{(j, k)}\otimes\bar{\rho}_{E}^{(k)}\, .
\end{equation*}
Since $k$ can be considered as a function of $j$, we can define $\rho_{E}^{(j)}\equiv\bar{\rho}_{E}^{(k(j))}$ and rewrite the above equation in a simpler form 
\begin{equation}
\label{eq:forty one}
\begin{aligned}
\rho_{ABE}=\bigoplus_{j}p_{j}\,\rho_{a_{j}^{L}B}\otimes\rho_{a_{j}^{R}}\otimes\rho_{E}^{(j)} \\
=\bigoplus_{j}p_{j}\,\rho_{A_{j}B}\otimes\rho_{E}^{(j)}\, ,\qquad
\end{aligned}
\end{equation}
where $\rho_{A_{j}B}\equiv\rho_{a_{j}^{L}B}\otimes\rho_{a_{j}^{R}}$ and, in addition, we have omitted the superscript $(j, k)$ of $\rho_{a_{j}^{R}}^{(j, k)}$. 

Similarly, Using Eq.~(\ref{eq:forty}), we can rewrite Eq.~(\ref{eq:twenty three}) as
\begin{equation}
\begin{aligned}
\label{eq:forty two}
\rho_{ABE}=\bigoplus_{k}q_{k}\,\rho_{A\,b_{k}^{L}}\otimes\rho_{b_{k}^{R}}^{(j, k)}\otimes\hat{\bar{\rho}}_{E}^{(j)}\, \\
=\bigoplus_{k}q_{k}\,\rho_{A\,b_{k}^{L}}\otimes\rho_{b_{k}^{R}}\otimes\hat{\rho}_{E}^{(k)} \;\; \\
=\bigoplus_{k}q_{k}\,\hat{\rho}_{AB_{k}}\otimes\hat{\rho}_{E}^{(k)}\, ,\qquad\quad
\end{aligned}
\end{equation}
where $\hat{\rho}_{E}^{(k)}\equiv\hat{\bar{\rho}}_{E}^{(j(k))}$ and $\hat{\rho}_{AB_{k}}\equiv\rho_{A\,b_{k}^{L}}\otimes\rho_{b_{k}^{R}}=\rho_{A\,b_{k}^{L}}\otimes\rho_{b_{k}^{R}}^{(j, k)}$.

In summary, each localized dynamics, as $id_{A}\otimes \mathcal{F}_{BE}$ or $id_{B}\otimes \mathcal{F}_{AE}$, reduces to a localized subdynamics, as $id_{A}\otimes\bar{\mathcal{E}}_{B}$ or $id_{B}\otimes\bar{\mathcal{E}}_{A}$, respectively, if and only if both Eqs.~(\ref{eq:forty one}) and (\ref{eq:forty two}) hold for the initial $\rho_{ABE}$. Note that in Eq.~(\ref{eq:twenty three}), $A$ and $E$ are separated (i.e. $\rho_{AE}=\mathrm{Tr_{B}}(\rho_{ABE})$ is separable), and in Eq.~(\ref{eq:thirty five}), $B$ and $E$ are separated. Now, the requirement that the both Eqs.~(\ref{eq:twenty three}) and (\ref{eq:thirty five}) must be held simultaneously, leads to Eqs.~(\ref{eq:forty one}) and (\ref{eq:forty two}), which in both, $E$ is separated from the whole $S=AB$. 

Let's end this section by considering an special (maybe interesting) case. Assume that for $j=j_{0}$ and all $k$, $\sigma_{a_{j_{0}}^{L}B_{k}}$ and $\sigma_{A_{j_{0}}b_{k}^{L}}$ in Eq.~(\ref{eq:thirty seven}) are non-zero. So, Eq.~(\ref{eq:thirty nine}) holds for this fixed $j=j_{0}$ and all $k$:
\begin{equation*}
\rho_{a_{j_{0}}^{R}\,E}=\rho_{a_{j_{0}}^{R}}^{(j_{0}, k)}\otimes\bar{\rho}_{E}^{(k)}.
\end{equation*}
Tracing from the both sides with respect to $a_{j_{0}}^{R}$, we get $\bar{\rho}_{E}^{(k)}=\mathrm{Tr_{\mathit{a_{j_{0}}^{R}}}}(\rho_{a_{j_{0}}^{R}\,E})$, which is a fixed state for all $k$. So all $\rho_{E}^{(j)}=\bar{\rho}_{E}^{(k)}$ in Eq.~(\ref{eq:forty one}) are the same (which we may denote it as $\rho_{E}$). Therefore 
\begin{equation*}
\rho_{ABE}=(\bigoplus_{j}p_{j}\,\rho_{a_{j}^{L}B}\otimes\rho_{a_{j}^{R}})\otimes\rho_{E}\, ;
\end{equation*}
i.e. the initial $\rho_{ABE}$ is factorized.

\section{When each part of the system interacts with its local environment}\label{sec:individual environment}
Now, let's consider the following case which may be more interesting than the previous one. Assume that the two parties $A$ and $B$ of our bipartite system, are separated from each other and each one interacts with its own local environment. Let's denote the local environment of $A$ as $E_{A}$, the local environment of $B$ as $E_{B}$ and the initial state of the system-environments as $\rho_{AE_{A}BE_{B}}$. From Sec.~\ref{sec:localized subdynamics}, we know that if, for a $\rho_{AE_{A}BE_{B}}$, each localized dynamics $id_{AE_{A}}\otimes \mathcal{F}_{BE_{B}}$ as 
\begin{equation}
\label{eq:forty three}
\begin{aligned}
\rho_{AE_{A}B^{\prime}E^{\prime}_{B}}^{\prime}=\sum_{j}(I_{AE_{A}}\otimes f_{j})\rho_{AE_{A}BE_{B}}(I_{AE_{A}}\otimes f_{j}^{\dagger})\, , \\
f_{j}:\,\cH_{B}\otimes\cH_{E_{B}}\rightarrow\,\cH_{B^{\prime}}\otimes\cH_{E_{B}^{\prime}}\, , \qquad\qquad\\
\sum_{j}f_{j}^{\dagger}\,f_{j}=I_{BE_{B}}\, , \qquad\qquad\qquad\quad
\end{aligned}
\end{equation}
reduces to a localized subdynamics $id_{AE_{A}}\otimes \bar{\mathcal{E}}_{B}$ as 
\begin{equation}
\label{eq:forty four}
\begin{aligned}
\rho_{AE_{A}B^{\prime}}^{\prime}=\sum_{i}(I_{AE_{A}}\otimes \bar{E_{i}})\rho_{AE_{A}B}(I_{AE_{A}}\otimes \bar{E_{i}}^{\dagger})\, , \\
\bar{E_{i}}:\,\cH_{B}\rightarrow\,\cH_{B^{\prime}}\, , \qquad\qquad\qquad\\
\sum_{i}\bar{E_{i}}^{\dagger}\bar{E_{i}}=I_{B}\, , \qquad\qquad\qquad\quad
\end{aligned}
\end{equation}
then the initial $\rho_{AE_{A}BE_{B}}$ is as
\begin{equation}
\label{eq:forty five}
\begin{aligned}
\rho_{AE_{A}BE_{B}}=\bigoplus_{k}q_{k}\,\rho_{AE_{A}\,b_{k}^{L}}\otimes\rho_{b_{k}^{R}E_{B}} ,\, \\
\cH_{B}=\bigoplus_{k}\cH_{B_{k}}=\bigoplus_{k}\cH_{b_{k}^{L}}\otimes\cH_{b_{k}^{R}}\, ,
\end{aligned}
\end{equation}
and vice versa. In above equations, $\rho_{AE_{A}B}=\mathrm{Tr_{E_{B}}}(\rho_{AE_{A}BE_{B}})$, $\rho_{AE_{A}B^{\prime}}^{\prime}=\mathrm{Tr_{E_{B}^{\prime}}}(\rho_{AE_{A}B^{\prime}E_{B}^{\prime}}^{\prime})$ and the final Hilbert spaces of $B$ and $E_{B}$, $\cH_{B^{\prime}}$ and $\cH_{E_{B}^{\prime}}$, may differ from the initial $\cH_{B}$ and $\cH_{E_{B}}$, respectively. 
In addition, $\lbrace q_{k}\rbrace$ is a probability distribution, $\rho_{AE_{A}b_{k}^{L}}$ is a state on $\cH_{A}\otimes\cH_{E_{A}}\otimes\cH_{b_{k}^{L}}$ and $\rho_{b_{k}^{R}E_{B}}$ is a state on $\cH_{b_{k}^{R}}\otimes\cH_{E_{B}}$. Also note that from Eq.~(\ref{eq:forty four}), we have
\begin{equation}
\label{eq:forty six}
\begin{aligned}
\rho_{AB^{\prime}}^{\prime}=\mathrm{Tr_{E_{A}}}(\rho_{AE_{A}B^{\prime}}^{\prime})=\sum_{i}(I_{A}\otimes \bar{E_{i}})\rho_{AB}(I_{A}\otimes \bar{E_{i}}^{\dagger}) \\
=id_{A}\otimes\bar{\mathcal{E}}_{B}(\rho_{AB})\, ,\qquad\qquad\qquad\quad
\end{aligned}
\end{equation}
where $\rho_{AB}=\mathrm{Tr_{E_{A}}}(\rho_{AE_{A}B})$ is the initial state of the system.

Similarly, if, for an initial $\rho_{AE_{A}BE_{B}}$, any arbitrary localized dynamics as $\rho_{A^{\prime}E_{A}^{\prime}BE_{B}}^{\prime}=\mathcal{F}_{AE_{A}}\otimes id_{BE_{B}}(\rho_{AE_{A}BE_{B}})$ reduces to a localized subdynamics as $\rho_{A^{\prime}BE_{B}}^{\prime}=\bar{\mathcal{E}}_{A}\otimes id_{BE_{B}}(\rho_{ABE_{B}})$, where $\rho_{ABE_{B}}=\mathrm{Tr_{E_{A}}}(\rho_{AE_{A}BE_{B}})$ and $\rho_{A^{\prime}BE_{B}}^{\prime}=\mathrm{Tr_{E_{A}^{\prime}}}(\rho_{A^{\prime}E_{A}^{\prime}BE_{B}}^{\prime})$, then
\begin{equation}
\label{eq:forty seven}
\begin{aligned}
\rho_{AE_{A}BE_{B}}=\bigoplus_{j}p_{j}\,\rho_{a_{j}^{L}E_{A}}\otimes\rho_{a_{j}^{R}BE_{B}},\, \\
\cH_{A}=\bigoplus_{j}\cH_{A_{j}}=\bigoplus_{j}\cH_{a_{j}^{L}}\otimes\cH_{a_{j}^{R}}\, ,
\end{aligned}
\end{equation}
and vice versa. In the above equation, $\lbrace p_{j}\rbrace$ is a probability distribution, $\rho_{a_{j}^{L}E_{A}}$ is a state on $\cH_{a_{j}^{L}}\otimes\cH_{E_{A}}$ and $\rho_{a_{j}^{R}BE_{B}}$ is a state on $\cH_{a_{j}^{R}}\otimes\cH_{B}\otimes\cH_{E_{B}}$.


Now, if, for an initial $\rho_{AE_{A}BE_{B}}$, each localized dynamics as $id_{AE_{A}}\otimes \mathcal{F}_{BE_{B}}$ or $\mathcal{F}_{AE_{A}}\otimes id_{BE_{B}}$ reduces to a localized subdynamics as $id_{AE_{A}}\otimes \bar{\mathcal{E}}_{B}$ or $\bar{\mathcal{E}}_{A}\otimes id_{BE_{B}}$, respectively, then both Eqs.~(\ref{eq:forty five}) and (\ref{eq:forty seven}) hold simultaneously (and vice versa).

Define the projection
\begin{equation}
\label{eq:forty eight}
\Pi_{k}=\Pi_{B_{k}}\otimes I_{AE_{A}E_{B}},
\end{equation}
where $\Pi_{B_{k}}$ is the projection onto $\cH_{B_{k}}$ and $I_{AE_{A}E_{B}}$ is the identity operator on $\cH_{A}\otimes\cH_{E_{A}}\otimes\cH_{E_{B}}$. So, using Eqs.~(\ref{eq:forty five}), (\ref{eq:forty seven}) and (\ref{eq:forty eight}), we have
\begin{equation}
\label{eq:forty nine}
\begin{aligned}
\Pi_{k}\,\rho_{AE_{A}BE_{B}}\,\Pi_{k}=q_{k}\,\rho_{AE_{A}\,b_{k}^{L}}\otimes\rho_{b_{k}^{R}E_{B}}\,\;\qquad \\
=\bigoplus_{j}p_{j}\,\rho_{a_{j}^{L}E_{A}}\otimes\sigma_{a_{j}^{R}B_{k}E_{B}}\, ,
\end{aligned}
\end{equation}
where $\sigma_{a_{j}^{R}B_{k}E_{B}}=\bar{\Pi}_{jk}\,\rho_{a_{j}^{R}BE_{B}}\,\bar{\Pi}_{jk}$ and $\bar{\Pi}_{jk}\equiv\Pi_{a_{j}^{R}}\otimes\Pi_{B_{k}}\otimes I_{E_{B}}$ (where $\Pi_{a_{j}^{R}}$ is the projection onto $\cH_{a_{j}^{R}}$). $\sigma_{a_{j}^{R}B_{k}E_{B}}$ is a positive operator on $\cH_{a_{j}^{R}}\otimes\cH_{B_{k}}\otimes\cH_{E_{B}}$. Let $p^{\prime}_{jk}=\mathrm{Tr}(\sigma_{a_{j}^{R}B_{k}E_{B}})$; so $0\leq p^{\prime}_{jk}\leq 1$. Now if $p^{\prime}_{jk}>0$, we define
\begin{equation*}
\rho_{a_{j}^{R}B_{k}E_{B}}=\dfrac{\sigma_{a_{j}^{R}B_{k}E_{B}}}{p^{\prime}_{jk}} ,
\end{equation*}
otherwise, if $p^{\prime}_{jk}=0$, we define $\rho_{a_{j}^{R}B_{k}E_{B}}$ arbitrarily.
So, Eq.~(\ref{eq:forty nine}) can be rewritten as
\begin{equation*}
q_{k}\,\rho_{AE_{A}b_{k}^{L}}\otimes\rho_{b_{k}^{R}E_{B}}
=\bigoplus_{j}p_{j}p^{\prime}_{jk}\,\rho_{a_{j}^{L}E_{A}}\otimes\rho_{a_{j}^{R}B_{k}E_{B}} .\,
\end{equation*}
Tracing from both sides, with respect to $b_{k}^{R}$ and $E_{B}$, we get
\begin{equation}
\label{eq:fifty}
\rho_{AE_{A}b_{k}^{L}}
=\bigoplus_{j}p_{jk}\,\rho_{a_{j}^{L}E_{A}}\otimes\rho_{a_{j}^{R}b_{k}^{L}}\, ,
\end{equation}
where $\rho_{a_{j}^{R}b_{k}^{L}}=\mathrm{Tr_{b_{k}^{R}E_{B}}}(\rho_{a_{j}^{R}B_{k}E_{B}})$ and $p_{jk}=\dfrac{p_{j}p^{\prime}_{jk}}{q_{k}}$ (note that we only consider those terms in Eqs.~(\ref{eq:forty five}) and (\ref{eq:forty seven}) for which $q_{k}\neq 0$ and $p_{j}\neq 0$). So we can rewrite  Eqs.~(\ref{eq:forty five}) as
\begin{equation}
\label{eq:fifty one}
\begin{aligned}
\rho_{AE_{A}BE_{B}}=\bigoplus_{jk}q_{k}p_{jk}\,\rho_{a_{j}^{L}E_{A}}\otimes\rho_{a_{j}^{R}b_{k}^{L}}\otimes\rho_{b_{k}^{R}E_{B}} \\
=\bigoplus_{jk}q_{jk}\,\rho_{a_{j}^{L}E_{A}}\otimes\rho_{a_{j}^{R}b_{k}^{L}}\otimes\rho_{b_{k}^{R}E_{B}}\; ,\\
\cH_{A}=\bigoplus_{j}\cH_{a_{j}^{L}}\otimes\cH_{a_{j}^{R}}\; ,\cH_{B}=\bigoplus_{k}\cH_{b_{k}^{L}}\otimes\cH_{b_{k}^{R}}\, ,
\end{aligned}
\end{equation}
where $q_{jk}=q_{k}p_{jk}$. Note that $\rho_{AE_{B}}=\mathrm{Tr_{BE_{A}}}(\rho_{AE_{A}BE_{B}})$, $\rho_{BE_{A}}=\mathrm{Tr_{AE_{B}}}(\rho_{AE_{A}BE_{B}})$ and $\rho_{E_{A}E_{B}}=\mathrm{Tr_{AB}}(\rho_{AE_{A}BE_{B}})$ are all separable states, but $\rho_{AB}$ my be entangled.

In addition,
\begin{equation}
\label{eq:fifty two}
\begin{aligned}
\rho_{AB}=\mathrm{Tr_{E_{A}E_{B}}}(\rho_{AE_{A}BE_{B}})\qquad\qquad\quad \\
=\bigoplus_{jk}q_{jk}\,\rho_{a_{j}^{L}}\otimes\rho_{a_{j}^{R}b_{k}^{L}}\otimes\rho_{b_{k}^{R}} ,\;
\end{aligned}
\end{equation}
where $\rho_{a_{j}^{L}}=\mathrm{Tr_{E_{A}}}(\rho_{a_{j}^{L}E_{A}})$ and $\rho_{b_{k}^{R}}=\mathrm{Tr_{E_{B}}}(\rho_{b_{k}^{R}E_{B}})$. So, if  (e.g., using the method introduced before Eq.~(\ref{eq:twenty two})) we construct the CP assignment maps $\Lambda_{a_{j}^{L}}:\mathcal{L}(\cH_{a_{j}^{L}})\rightarrow\mathcal{L}(\cH_{a_{j}^{L}}\otimes\cH_{E_{A}})$ and $\Lambda_{b_{k}^{R}}:\mathcal{L}(\cH_{b_{k}^{R}})\rightarrow\mathcal{L}(\cH_{b_{k}^{R}}\otimes\cH_{E_{B}})$ as 
\begin{equation}
\label{eq:fifty three}
\begin{aligned}
\Lambda_{a_{j}^{L}}(\rho_{a_{j}^{L}})=\rho_{a_{j}^{L}E_{A}}\;, \\
\Lambda_{b_{k}^{R}}(\rho_{b_{k}^{R}})=\rho_{b_{k}^{R}E_{B}}\;,
\end{aligned}
\end{equation}
then we can define the CP assignment map $\Lambda:\mathcal{L}(\cH_{A}\otimes\cH_{B})\rightarrow\mathcal{L}(\cH_{A}\otimes\cH_{E_{A}}\otimes\cH_{B}\otimes\cH_{E_{B}})$ as $\rho_{AE_{A}BE_{B}}=\Lambda(\rho_{AB})=\Lambda_{A}\otimes\Lambda_{B}(\rho_{AB})$, where $\Lambda_{A}=\bigoplus_{j}\Lambda_{a_{j}^{L}}\otimes id_{a_{j}^{R}}$ and $\Lambda_{B}=\bigoplus_{k}id_{b_{k}^{L}}\otimes\Lambda_{b_{k}^{R}}$ are CP assignment maps on $\mathcal{L}(\cH_{A})$ and $\mathcal{L}(\cH_{B})$, respectively ($id_{a_{j}^{R}}$ and $id_{b_{k}^{L}}$ are the identity maps on $\mathcal{L}(\cH_{a_{j}^{R}})$ and $\mathcal{L}(\cH_{b_{k}^{L}})$, respectively).  

Therefore, for each localized map as $\mathcal{F}_{AE_{A}}\otimes \mathcal{F}_{BE_{B}}$, where $\mathcal{F}_{AE_{A}}$ ($\mathcal{F}_{BE_{B}}$) is a CP map on $\mathcal{L}(\cH_{A}\otimes\cH_{E_{A}})$ ($\mathcal{L}(\cH_{B}\otimes\cH_{E_{B}}$)), we have
\begin{equation}
\label{eq:fifty four}
\begin{aligned}
\rho^{\prime}_{AB}=\mathrm{Tr_{E_{A}E_{B}}}(\rho^{\prime}_{AE_{A}BE_{B}})\qquad\qquad\qquad\qquad\qquad \qquad\\
=\mathrm{Tr_{E_{A}E_{B}}}[\mathcal{F}_{AE_{A}}\otimes \mathcal{F}_{BE_{B}}(\rho_{AE_{A}BE_{B}})] \qquad\qquad\qquad \quad \\
=\mathrm{Tr_{E_{A}E_{B}}}[(\mathcal{F}_{AE_{A}}\otimes \mathcal{F}_{BE_{B}})(\Lambda_{A}\otimes\Lambda_{B})(\rho_{AB})] \qquad\quad \\ 
=(\mathrm{Tr_{E_{A}}}\circ \mathcal{F}_{AE_{A}}\circ\Lambda_{A})\otimes(\mathrm{Tr_{E_{B}}}\circ \mathcal{F}_{BE_{B}}\circ\Lambda_{B})(\rho_{AB}) \\
=\mathcal{\bar{E}}_{A}\otimes \mathcal{\bar{E}}_{B}\,(\rho_{AB}),\qquad\qquad\qquad\qquad\qquad\qquad
\end{aligned}
\end{equation}
where $\mathcal{\bar{E}}_{A}\equiv\mathrm{Tr_{E_{A}}}\circ \mathcal{F}_{AE_{A}}\circ\Lambda_{A}$ and $\mathcal{\bar{E}}_{B}\equiv\mathrm{Tr_{E_{B}}}\circ \mathcal{F}_{BE_{B}}\circ\Lambda_{B}$ are CP maps on $\mathcal{L}(\cH_{A})$ and $\mathcal{L}(\cH_{B})$, respectively.

In summary, if the initial $\rho_{AE_{A}BE_{B}}$ be as Eq.~(\ref{eq:fifty one}), then each localized dynamics as $\mathcal{F}_{AE_{A}}\otimes \mathcal{F}_{BE_{B}}$ reduces to a localized subdynamics as $\mathcal{\bar{E}}_{A}\otimes \mathcal{\bar{E}}_{B}$. In addition, if each localized dynamics as $id_{AE_{A}}\otimes \mathcal{F}_{BE_{B}}$ or $\mathcal{F}_{AE_{A}}\otimes id_{BE_{B}}$ reduces to a localized subdynamics as $id_{AE_{A}}\otimes \bar{\mathcal{E}}_{B}$ or $\bar{\mathcal{E}}_{A}\otimes id_{BE_{B}}$, respectively, then the initial $\rho_{AE_{A}BE_{B}}$ is given by Eq.~(\ref{eq:fifty one}) (and vice versa).

\section{Conclusion}\label{sec:summary}
We considered a bipartite quantum system including parties $A$ and $B$. Assuming that the dynamics of the system-environment is given by $id_{A}\otimes \mathcal{F}_{BE}$, we questioned whether the reduced dynamics of the system is as $id_{A}\otimes \bar{\mathcal{E}}_{B}$. At the first look, one may expect that this will be the case, since when the dynamics is given by $id_{A}\otimes \mathcal{F}_{BE}$, it means that the part $A$ is isolated from the environment and its reduced state remains unchanged during the evolution. But, as we saw in Sec.~\ref{sec:localized subdynamics}, only for Markov states, Eq.~(\ref{eq:twenty three}), each dynamics as $id_{A}\otimes \mathcal{F}_{BE}$ reduces to a subdynamics as $id_{A}\otimes \bar{\mathcal{E}}_{B}$.
In addition, we gave some illustrating examples in that section, too.
%

In Secs.~\ref{sec:markov}, we proved that the initial $\rho_{ABE}$ is a Markov state if and only if 
 each localized dynamics as Eq.~(\ref{eq:B1}) \textit{directly} reduces to a localized subdynamics. When the initial $\rho_{ABE}$ is not a Markov state, then Theorem 1 states that one can find, at least, one localized dynamics which its reduction is not equivalent to any localized subdynamics. Whether this localized dynamics is in the form of Eq.~(\ref{eq:B1}) or in the (more general) form of Eq.~(\ref{eq:twenty five}), remained as an open question.

In Secs.~\ref{sec:both part interaction} and \ref{sec:individual environment}, we generalized the  result given in Sec.~\ref{sec:localized subdynamics}. When the both parts of the system, $A$ and $B$, can interact with the environment was considered in  Sec.~\ref{sec:both part interaction} and when each part of the system interacts with its local environment was discussed in Sec.  \ref{sec:individual environment}. For example, in Sec.~\ref{sec:individual environment}, we have shown that each localized dynamics as $\mathcal{F}_{AE_{A}}\otimes \mathcal{F}_{BE_{B}}$, where $E_{A}$($E_{B}$) is the local environment of $A$ ($B$), reduces to a localized subdynamics as $\mathcal{\bar{E}}_{A}\otimes \mathcal{\bar{E}}_{B}$, if the initial state of the system and its environments is given by Eq.~(\ref{eq:fifty one}).

\section*{Acknowledgments}

We would like to thank the anonymous referee for his/her helpful comments.


\end{document}